\documentclass[pra,floatfix,nofootinbib,superscriptaddress,notitlepage, twocolumn, amsmath,amssymb,amsfonts]{revtex4-1} 

\usepackage[T1]{fontenc}
\usepackage{lmodern} 
\usepackage{xcolor}
\usepackage{bbold}
\usepackage{dsfont}
\usepackage{amsmath}
\usepackage{graphicx}
\usepackage{svg}
\usepackage[caption=false]{subfig}
\usepackage[colorlinks]{hyperref}
\hypersetup{
  colorlinks,
  citecolor=blue,
  linkcolor=blue,
  urlcolor=blue}
\usepackage[bottom]{footmisc}
\usepackage{enumerate}
\usepackage{float}
\usepackage{mathtools}
\usepackage[draft,inline,nomargin]{fixme}
\usepackage[normalem]{ulem}
\usepackage{balance}

\usepackage{diagbox}

\DeclarePairedDelimiter\bra{\langle}{\rvert}
\DeclarePairedDelimiter\ket{\lvert}{\rangle}

\DeclarePairedDelimiterX\braket[2]{\langle}{\rangle}{#1 \delimsize\vert #2}
\DeclarePairedDelimiterX\ketbra[2]{\vert}{\lvert}{#1 \delimsize\rangle\langle #2}

\begin{document}
\newcommand{\todo}[1]{\red{{$\bigstar$\sc #1}$\bigstar$}}
\newcommand{\ks}[1]{{\textcolor{teal}{[KS: #1]}}}
\global\long\def\eqn#1{\begin{align}#1\end{align}}
\global\long\def\ket#1{\left|#1\right\rangle }
\global\long\def\bra#1{\left\langle #1\right|}
\global\long\def\bkt#1{\left(#1\right)}
\global\long\def\sbkt#1{\left[#1\right]}
\global\long\def\cbkt#1{\left\{#1\right\}}
\global\long\def\abs#1{\left\vert#1\right\vert}
\global\long\def\der#1#2{\frac{{d}#1}{{d}#2}}
\global\long\def\pard#1#2{\frac{{\partial}#1}{{\partial}#2}}
\global\long\def\re{\mathrm{Re}}
\global\long\def\im{\mathrm{Im}}
\global\long\def\dd{\mathrm{d}}
\global\long\def\ddd{\mathcal{D}}

\global\long\def\avg#1{\left\langle #1 \right\rangle}
\global\long\def\mr#1{\mathrm{#1}}
\global\long\def\mb#1{{\mathbf #1}}
\global\long\def\mc#1{\mathcal{#1}}
\global\long\def\tr{\mathrm{Tr}}
\global\long\def\dbar#1{\Bar{\Bar{#1}}}

\global\long\def\nth{$n^{\mathrm{th}}$\,}
\global\long\def\mth{$m^{\mathrm{th}}$\,}
\global\long\def\non{\nonumber}
\newcommand{\teal}[1]{{\color{teal} {#1}}}

\newcommand{\orange}[1]{{\color{orange} {#1}}}
\newcommand{\cyan}[1]{{\color{cyan} {#1}}}
\newcommand{\blue}[1]{{\color{blue} {#1}}}
\newcommand{\yellow}[1]{{\color{yellow} {#1}}}
\newcommand{\green}[1]{{\color{green} {#1}}}
\newcommand{\red}[1]{{\color{red} {#1}}}
\global\long\def\todo#1{\yellow{{$\bigstar$ \orange{\bf\sc #1}}$\bigstar$} }

\title{Nanofiber-based second-order atomic Bragg lattice for collectively enhanced coupling} 
\author{N. Vera}
\email{nicvera@udec.cl }
\affiliation{Departamento de F\'isica, Facultad de Ciencias F\'isicas y Matem\'aticas, Universidad de Concepci\'on, Concepci\'on, Chile}

\author{P. Solano}
\email{psolano@udec.cl}
\affiliation{Departamento de F\'isica, Facultad de Ciencias F\'isicas y Matem\'aticas, Universidad de Concepci\'on, Concepci\'on, Chile}

\begin{abstract}

We propose two experimental schemes for nanofiber-based compensated optical dipole traps that optimize the collective coupling of a one-dimensional array of atoms. The created array satisfies the second-order Bragg condition ($d=\lambda$), facilitating constructive interference of atomic radiation into the nanofiber and generating coherent back reflections of guided modes. Both schemes use far-off resonance light to minimize light scattering and atomic heating. Our numerical study focuses on $^{87}$Rb atoms. The results are generalizable to different atomic species and could improve the study of collective and nonlinear atomic effects.
\end{abstract}

\maketitle

\section{Introduction}
\label{subsec:sota_int}

The observation of collective atomic effects, such as superradiance, subradiance, and photonic band gaps \cite{deutsch1995photonic,dicke1954coherence,skribanowitz1973observation,gross1982superradiance}, benefits from ordered arrays, as experimentally demonstrated through atomic lattices interacting with resonant light \cite{birkl1995bragg,schilke2011photonic,rui2020subradiant,tamura2020phase}. These effects are enhanced when increasing the light-matter coupling to a preferential electromagnetic (EM) mode by spatial confinement, typically using resonators and focused Gaussian beams \cite{walther2006cavity}. Recently, nanoscale waveguides have been used to transversely confine EM modes through lengths beyond the limits of diffraction in free space, enhancing light-matter coupling for long chains of atoms, an enabling the possibility of connecting distant emitters \cite{sheremet2023waveguide,PhysRevApplied.13.064010}.

Tapered optical nanofibers (ONF) are an example of cylindrical symmetric, uniform-thickness nanoscale waveguides. Nearby atoms interact with the evanescent field of the guided optical mode, making an efficient and versatile platform for optical dipole traps and allowing large field-atom coupling over extended regions \cite{solano2017optical,Nieddu2016Jjop}. The figure of merit for resonant light-matter interactions is the coupling efficiency $\beta=\frac{\Gamma_{\rm{1D}}}{\Gamma_{\rm{1D}}+\Gamma_{\rm{out}}}$, where $\Gamma_{\rm{1D}}$ is the spontaneous decay rate of an atom into the guided EM field mode and $\Gamma_{\rm{1D}}+\Gamma_{\rm{out}}$ is
its total decay rate, meaning inside and outside the waveguide \cite{solano2017optical}. ONFs offer a coupling efficiency around $\beta\sim 0.01$ per optically trapped atom, strongly depending on the distance to the waveguide surface \cite{le2005spontaneous,PhysRevA.99.013822}. Regarding optical trapping, far-detuned laser beams counterpropagating through the ONF create a periodic optical potential, arranging the atoms into an effective 1-D lattice \cite{vetsch2010optical}.

\begin{figure}[t]
\centering
\includegraphics[width=0.45 \textwidth]{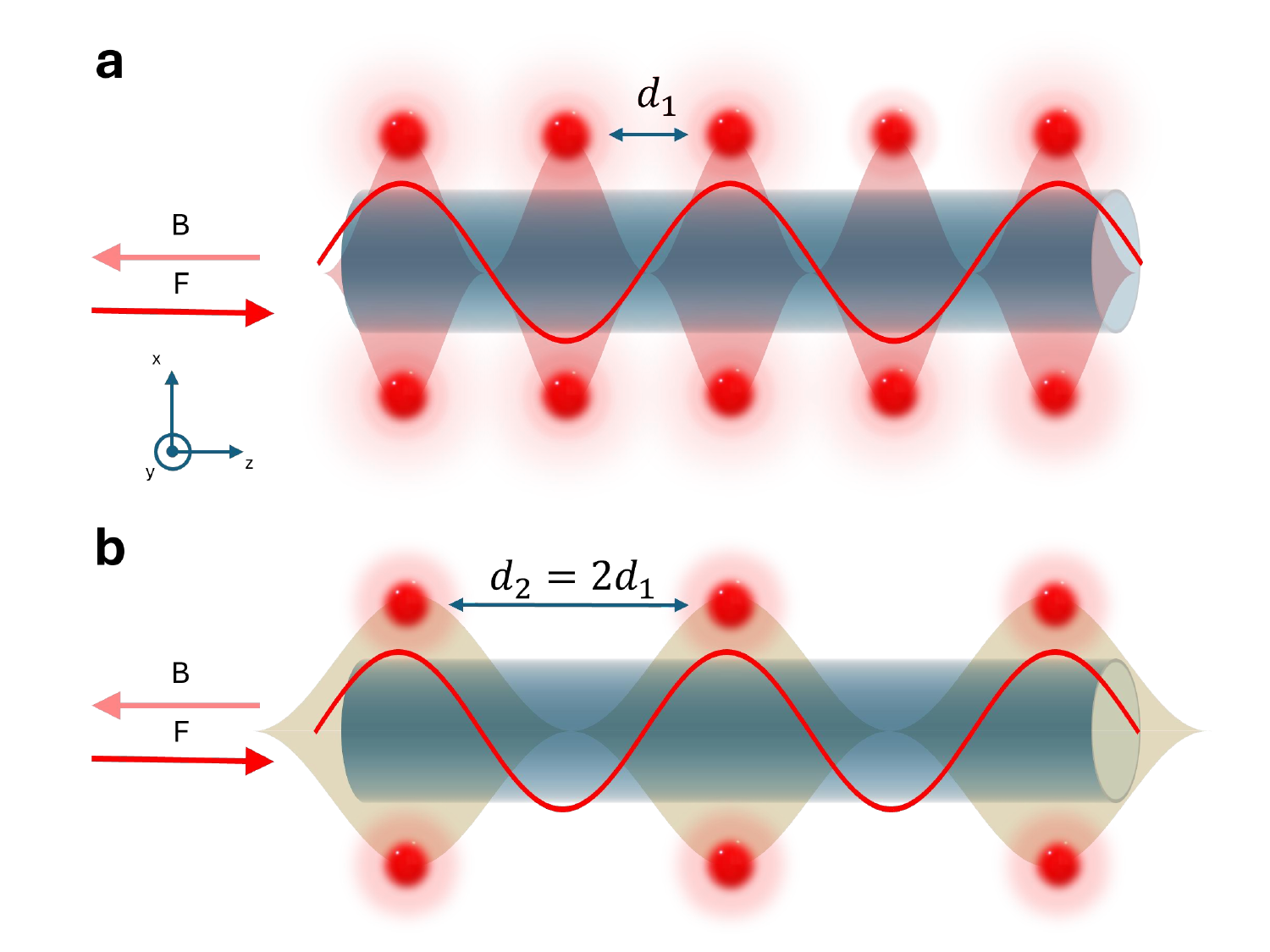}
\caption{\textbf{a.} The trapping standing wave (red shaded region) has a wavelength near atomic resonance, placing the atoms $\lambda / 2$ apart, but  suffering from large scattering rates and heating. \textbf{b.} The proposed trapping standing wave (yellow shaded region) arranges the atoms at twice the distance with far-off resonant light. Due to dispersion, this standing wave is not produced by twice the free space resonance wavelength.}
 \label{fig:scheme}
\end{figure}

Multiple atoms coupled to an ONF can radiate collectively, depending on their relative positions and initial conditions, leading to a collective coupling efficiency $\beta_c$. In the optimal case $\beta_c=\frac{N\Gamma_{\rm{1D}}}{N\Gamma_{\rm{1D}}+\Gamma_{\rm{out}}}$, which quickly approaches one as the number of atoms $N$ increases. Such an optimal collective coupling is achieved when the atoms are distanced by multiple integers of half their resonant wavelength \cite{chang2012cavity,cardenas2023many}. The lattice then forms a band gap for a guided probe if it satisfies the Bragg condition
\begin{equation}\label{eq:bragg_cond}
d_q = \frac{1}{2}q\lambda_{p},
\end{equation}
where $d_q$ is the distance between scatterers, $q$ is an integer for the diffraction order, and $\lambda_{p}$ is the probe wavelength within the ONF, which differs from the vacuum wavelength. Previous works \cite{corzo2016large,sorensen2016coherent} have demonstrated Bragg reflection in ONFs using a standing wave near resonance, corresponding to the first order ($q=1$), which leads to a highly uncompensated trap and a high photon scattering rate (see Fig. \ref{fig:scheme}). 

In this work, we propose to use the second-order condition, $q=2$, thus avoiding photon scattering and light shift due to the standing wave. This configuration adds other experimental advantages, such as easily separating the probe from the lattice beams and allowing the coupling of an external (non-guided) mode into the fiber through collectively enhanced scattering with the atomic ensemble. Since the mode propagation constant depends on the ONF radius and index of refraction, we numerically find the wavelength for the lattice laser beam that satisfies the condition. We calculate the lattice conditions for the  D2 transition in $ ^{87}$Rb within a fused silica ONF and simulate dipole traps in a two-color magic wavelength setup and a three-color power-compensated setup. The theoretical analysis and trapping schemes are suitable for other transitions and atomic species, presenting a useful tool for studying collective effects in hybrid atomic and photonic systems.

\section{Optical dipole traps in ONFs}
\label{subsec:dip_trap}

Optical dipole traps are usually implemented using light far-detuned from atomic resonances, generating a trapping potential for the atoms through the AC Stark shift \cite{chu1986experimental,Westbrook1990}. The atomic dynamic polarizability determines if a given EM field would be attractive or repulsive. This effect can be calculated for alkali atoms using the total light shifts expression \cite{steck2024quantum},
\begin{multline}
\label{eq:starkshift}
    \Delta E\left(F,m_f ;\omega\right) = -\alpha^{(0)}\left(F;\omega\right)\left|\mathcal{E}^{(+)}\right|^2 \\ -\alpha^{(1)}\left(i\mathcal{E}^{(-)}\times\mathcal{E}^{(+)} \right)_0 \frac{m_f}{F} \\ -\alpha^{(2)}\left(F;\omega\right)\frac{3\left|\mathcal{E}^{(+)}_0\right|^2- \left|\mathcal{E}^{(+)}\right|^2}{2}\left(\frac{3m_F^2-F(F+1)}{F(2F-1)}\right),
\end{multline}
where $\mathcal{E}_0^{(+)}$ is the electric field component in the quantization axis direction, $\alpha^{(i)}$ ($i=0,1,2$) are the scalar, vector and tensor polarizabilities, which can be calculated from the dipole matrix elements. $F$ is the hyperfine state level for the atom, and $m_F$ is the atom hyperfine Zeeman level. Usually, for alkali atoms, wavelengths blue detuned to the D2 line are used as repulsive fields, and wavelengths longer than the D1 line (red detuned) are used as attractive fields.

\begin{figure}
\centering
\includegraphics[width=0.5\textwidth]{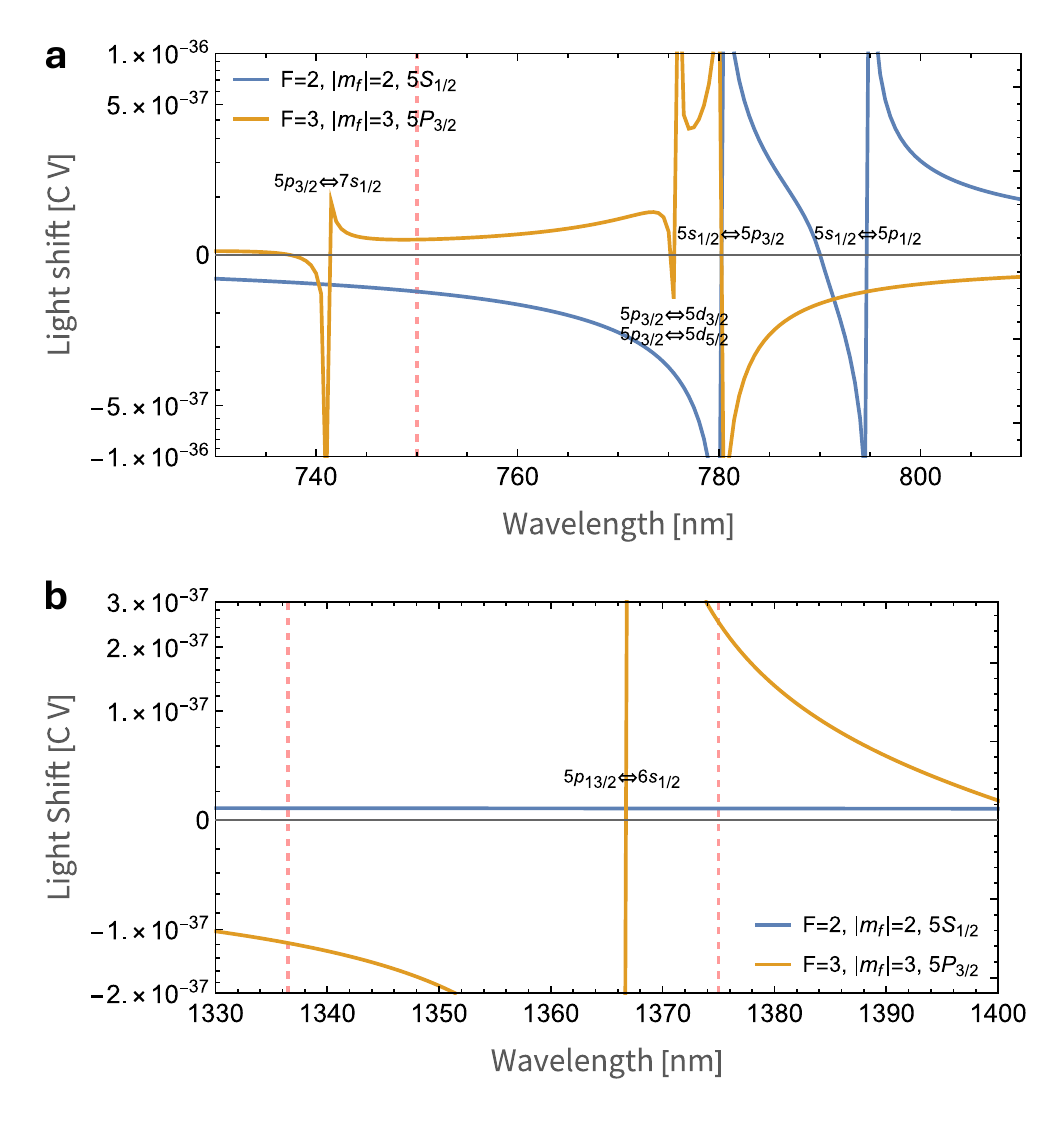}
\caption{Wavelength dependent unit amplitude atomic light shifts around \textbf{a} D1 and D2 lines and \textbf{b} the $5p_{3/2}\rightarrow6s_{1/2}$ transition. Negative (positive) values induce attractive (repulsive) potentials for the atoms. The vertical dashed red lines represent the selected wavelengths for the three color dipole trap.}
 \label{fig:polarizabilities}
\end{figure}

ONF-based optical dipole traps use the light-shifts produced by the evanescent field of the guided mode. Two counter-propagating beams generate a periodically spaced attractive potential for the atoms, providing radial and longitudinal confinement. A repulsive field is added to the attractive ones to prevent atoms from falling onto the ONF. The relative polarization between the attractive and repulsive fields can also provide azimuthal confinement \cite{vetsch2010optical,balykin2004atom,solano2017dynamics}. Van der Waals forces between the atoms and the fiber surface, which decays with the inverse cube of their distance, also play a role in the total potential at short distances \cite{frawley2012van,PhysRevA.97.032509}.

 Since the light shift (Eq. (\ref{eq:starkshift})) depends non trivially on the polarization structure of the electric field, we use a unitary amplitude electric field with an arbitrary polarization which we can match with the experimental conditions provided by the ONF, allowing us to compute magic wavelengths and zero crossings, to later compute the total shift for selected wavelengths \cite{ye2008quantum,goban2012demonstration,ding2012corrections,le2013dynamical}. Both trapping models we consider in this work, use a cross-polarization between the attractive, counter propagating lattice beam, and the running  repulse beam. In this circumstance, it is safe to ignore the light shifts due to the vector polarizability around the atom trapping sites . Fig. \ref{fig:polarizabilities} shows the unitary light shift for $^{87}$Rb for the ground and excited states around the D lines and in a far detuned region around 1370 nm.

\section{Lattice Condition}
\label{subsec:lat_cond}

To arrange the trapped atoms in the second-order Bragg condition (Eq. (\ref{eq:bragg_cond})), we must confine the atoms within a standing wave whose wavelength is twice the resonant one inside the ONF. In terms of the modes propagation constant, this reads
\begin{equation}\label{eq:bragg_cond2}
2k_{\mathrm{stand}}\left(a,\lambda\right)=k_{\mathrm{probe}}\left(a\right),
\end{equation}
where $a$ is the ONF radius, and $k_{\mathrm{stand}}$ and $k_{\mathrm{probe}}$ are the wavenumbers for the trapping and resonant probe light, respectively. The propagation constant for each mode is obtained from the fundamental (HE11) mode solution of the EM field boundary conditions inside the ONF. This solution comes from a numerically solved transcendental equation \cite{marcuse2013theory,hoffman2014ultrahigh}. 

\begin{figure}
\centering
\includegraphics[width=0.45 \textwidth]{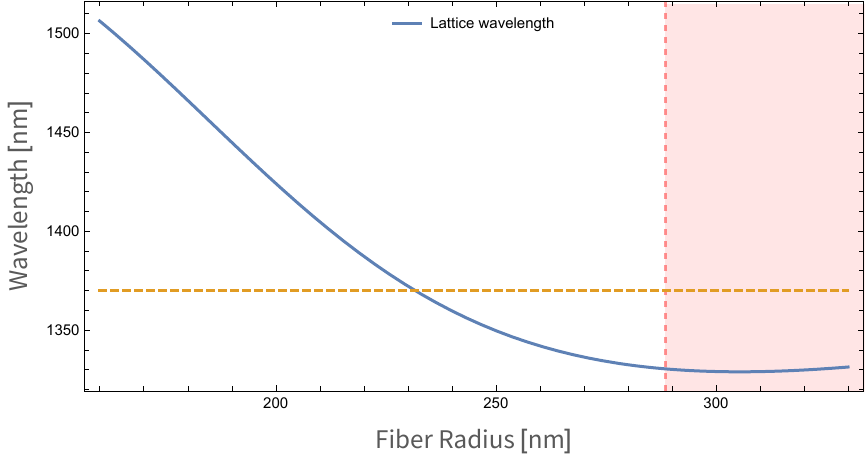}
\caption{Free space wavelength of the trapping laser that satisfies Eq. (\ref{eq:bragg_cond2}). The red-shaded area represents a forbidden zone where a repulsive field, blue-detuned from the D1 line, allows higher-order modes that distort the trapping potential. The orange dashed line is an excited state transition $5p_{3/2}\rightarrow6s_{1/2}$, shown in Fig. \ref{fig:polarizabilities}b, that has to be considered for the light-shift calculations of the excited state.}
 \label{fig:lat_cond}
\end{figure}

We calculated the Bragg condition (Eq. (\ref{eq:bragg_cond2})) using the Sellmeier coefficients for fused silica \cite{malitson1965interspecimen}, for the D2 transition of $ ^{87}$Rb \cite{steck2023rubidium}  (see Fig. \ref{fig:lat_cond}). This condition fixes the wavelength of the standing wave to the ONF radius.

To model the optical dipole trap, which satisfies Eq. (\ref{eq:bragg_cond2}) around the ONF, we use  Eq. (\ref{eq:starkshift}). We consider the $F_{g,2}$ $\rightarrow$ $F_{e,3}$ transition in $ ^{87}$Rb D2 line, with the atoms in the stretched states ($|m_{F_{g,2}}|=2$ and $|m_{F_{e,3}}|=3$). Bellow, we propose two schemes to achieve a nearly compensated ONF-based dipole trap that satisfies the second-order Bragg condition: a three-color power-compensated trap and a two-color magic wavelength trap. For the atomic scalar and tensor polarizabilites at the relevant wavelengths, we used the Alkali Rydberg Calculator (ARC) \cite{vsibalic2017arc} together with Nanotrappy \cite{berroir2022nanotrappy}. 

\subsubsection{Three color dipole trap}

\begin{figure}
\centering
\includegraphics[width=0.48 \textwidth]{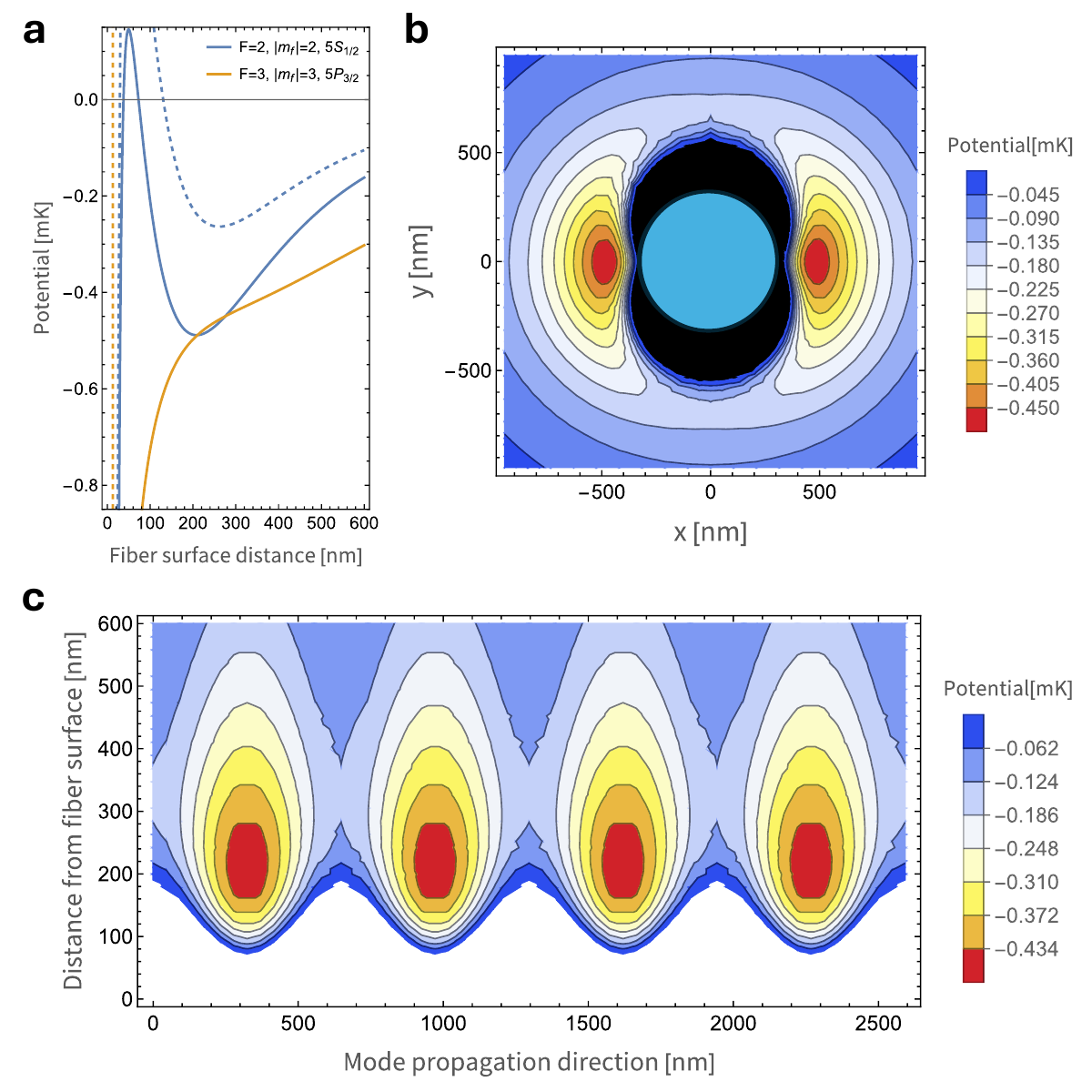}
\caption{\textbf{Potential profile around ONF for the three color dipole trap.} \textbf{a.} Potential as a function of distance from the fiber surface for the ground and excited levels, along the polarization of the lattice beam at an antinode. Continuous lines are obtained with 8 mW blue detuned laser, a 2$\times$2.39 mW compensating laser and 2$\times$2.4 mW lattice laser. Dashed lines are obtained turning off the compensating laser to show how it helps to compensate for the excited state shift. \textbf{b.} Ground state transverse profile at an anti-node position. \textbf{c.} Ground state longitudinal profile along the lattice beam polarization.}
 \label{fig:potential_profile}
\end{figure}

In the three color trap, we use a 750 nm, blue detuned repulsive field, and a 1336.25 nm standing wave, lattice beam (corresponding to a 270 nm fiber radius according to the calculated condition, see Fig. \ref{fig:lat_cond}). Unfortunately, the excited states are more light-shifted than the ground state in such conditions. A third compensating laser can largely mitigate such position-dependent resonance, significantly reducing inhomogeneous broadening and allowing continuous cooling of already trapped atoms. This is why we introduce a 1375 nm, power compensating laser, which mitigates the excited state shift due to the 1336.25 nm laser (See Fig. \ref{fig:polarizabilities}b). We use cross polarization between the lattice and blue detuned beams, whereas the 1375 nm polarization is parallel to the lattice beam. A pair of slightly detuned 1375 nm lasers are used to get rid of vector polarizability while averaging out the undesired secondary lattice. Fig. \ref{fig:potential_profile}a shows the potential profile for the ground and excited states before and after the 1375 nm compensating laser has been added, at an antinode of the lattice standing wave. The trap minimum, located around 210 nm from the ONF surface, determines a depth of 0.48 mK and a coupling efficiency of $\beta = 0.008$ per atom. Fig. \ref{fig:potential_profile}b and \ref{fig:potential_profile}c show the transverse and longitudinal potential profiles. This trap configuration has frequencies of $\{\omega_x,\omega_y,\omega_z \}/2\pi = \{225,147,209\}$ kHz for each potential well, which gives an atomic spread of $\{\sigma_x,\sigma_y,\sigma_z \} = \{23,24,28\}$ nm for atoms in the motional ground state \cite{fetter2012bose}.

\subsubsection{Magic wavelength dipole trap}

In this setup. we use a 626.65 nm running wave, blue-detuned repulsive field and 1407.84 nm standing wave (corresponding to a 208.32 nm radius ONF), lattice beam in cross polarization. Both wavelengths are magic wavelengths for the atom polarizabilities when taking into account the scalar and tensor polarizabilities in a quasi-linear polarization state of light for a quantization axis along the mode propagation direction. For the lattice beam, we determine the magic wavelength at the antinode. The magic wavelength does not hold elsewhere. Notice that a laser around 790 nm, between the D1 and D2 lines, could be used as a magic wavelength repulsive field as well.

It is worth noting that the lattice beam magic wavelength condition fixes the ONF radius to a particular value, not allowing for flexibility at the moment of manufacturing the ONF. In practice, it is hard to precisely manufacture an ONF with a radius near or under 200 nm, but an experimental realization in turn would not need to introduce a third power stabilized compensating beam, and there are ways of measuring ONF diameters non-destructively with high precision \cite{Fatemi:17}. Trap depth at its minima (around 210 nm) is 0.23 mK, with a coupling $\beta = 0.024$ per atom. Trapping frequencies are $\{\omega_x,\omega_y,\omega_z \}/2\pi = \{151,94,171\}$ kHz. When considering an atom in the potential well kinetic ground state, the atoms spread a distance $\{\sigma_x,\sigma_y,\sigma_z \} = \{27,26,35\}$ nm around the potential minima.

\begin{figure}
\centering
\includegraphics[width=0.48 \textwidth]{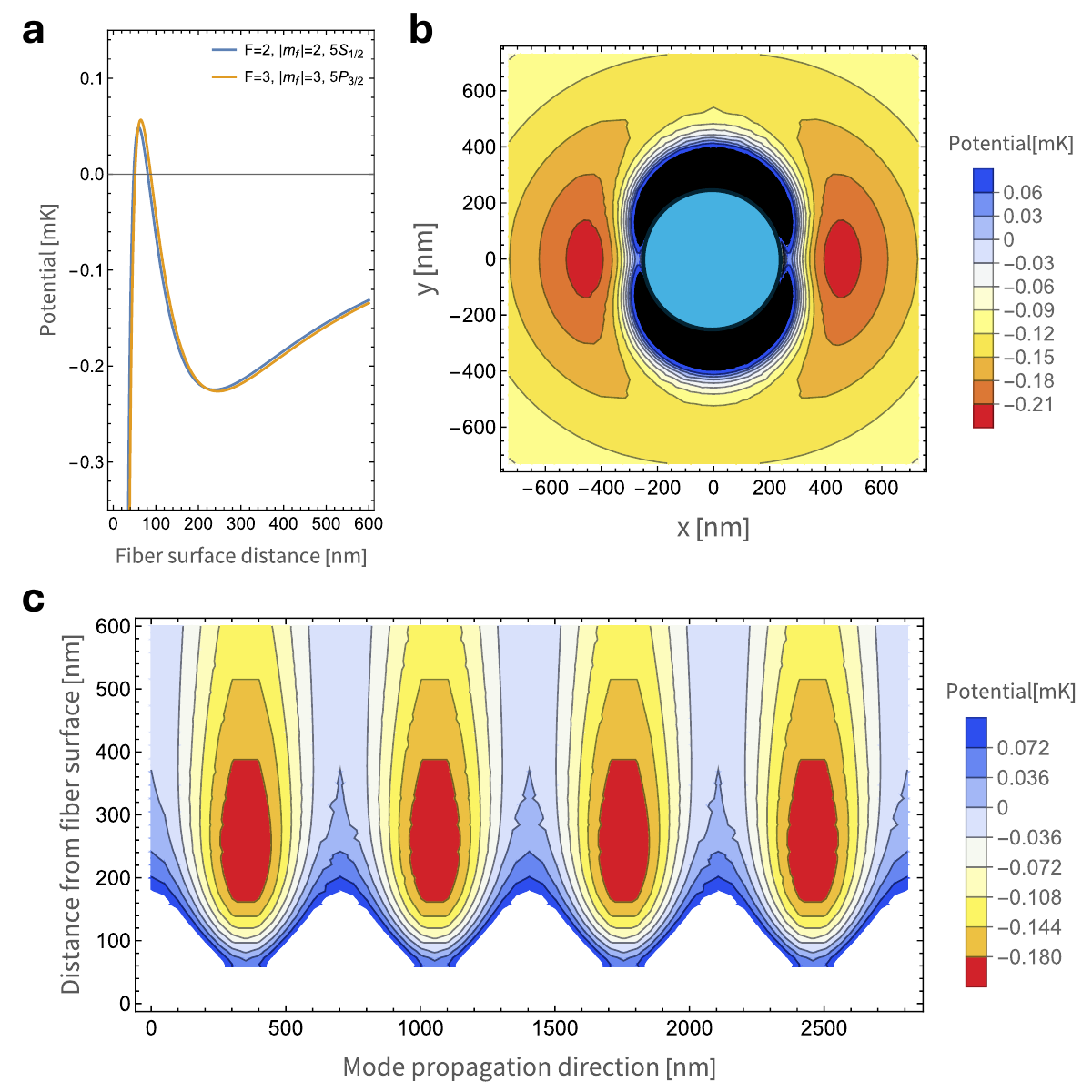}
\caption{\textbf{Potential profile around ONF for the magic wavelength dipole trap.} \textbf{a.} Potential as a function of distance from the fiber surface for the ground and excited levels, along the polarization of the lattice beam at an anti node. Continuous lines are obtained with 12 mW blue detuned laser and 2$\times$5 mW lattice laser. \textbf{b.} Transverse profile in an antinode position. \textbf{c.} Longitudinal profile along the lattice beam polarization.}
 \label{fig:potential_profile_mw}
\end{figure}

\section{Discussion}

While an external perpendicularly incident optical excitation results in a sub-radiant state in ensembles of atoms satisfying the first-order Bragg condition, $q=1$, the second-order Bragg condition, $q=2$, always leads to superradiance. The latter condition enhances the coupling of external excitations into the guided mode mediated by the atoms, guaranteeing that every optical excitation in the waveguide comes from a collective excitation of the atoms. For example, for two distant ensembles coupled via the same nanofiber, one can introduce excitations through one ensemble without exciting the other, guaranteeing interaction between ensembles without imposing amplitude or phase relation among them, as it happens with an otherwise guided excitation laser.\\

We have yet to address how the atomic spread within the dipole traps inhibits collective effects such as the Bragg reflection. Atomic random positions due to harmonic motion along the ONF deteriorate the interference visibility, decreasing Bragg's reflection and superradiance compared to the ideal case. For our proposed schemes, we predict a Debye-Waller factor close to 1, similar to Ref. \cite{corzo2016large}, making a negligible reduction of a Bragg reflection. Radial and azimuthal thermal motion could have a smaller effect on interference visibility, but a detailed analysis is beyond the scope of this work.\\

One might argue that since the condition in Eq. (\ref{eq:bragg_cond2}) implies that the number of atoms is halved in comparison to the near-resonance case, collective effects get deteriorated \cite{cardenas2023many}. However, since nanofibers allow for an arbitrarily long trapping region, this can easily be compensated for by loading more atoms along the ONF. Moreover, a higher potential well occupation and enhanced coupling would be expected due to the deep potentials, low scattering rate, and the possibility of using degenerate Raman cooling in the cross-polarized traps \cite{meng2018near}. 

\section{Conclusion}

In conclusion, we numerically find and characterize nanofiber-based optical dipole traps that optimize the collectively enhanced coupling between a cold atomic ensemble and a guided optical mode. The proposed schemes overcome most problems associated with previous implementations, which used near-resonant standing waves, such as high photon scattering rate, short trap lifetime, and challenges distinguishing between lattice and probe beams. Although we use $^{87}$Rb as an example, the second-order Bragg condition could be also implemented for other alkali atoms. The proposed experimental configuration has potential applications for implementing the Dicke condition for superradiance and superfluorescence, and test for mirror symmetry breaking in fully inverted ensembles. \\

This work is supported by the National Agency for Science and Technology (ANID), through Project FONDECYT Grants No. 1240204. The authors acknowledge Sebastien Garcia, Luis Orozco and Constanze Bach for insightful discussion.

\bibliography{ref}

\end{document}